\documentclass{PoS}

\usepackage{multirow}
\usepackage{subcaption}

\newcommand{\MSb}{\overline{\mathrm{MS}}}

\let\OLDthebibliography\thebibliography
\renewcommand\thebibliography[1]{
  \OLDthebibliography{#1}
  \setlength{\parskip}{0pt}
  \setlength{\itemsep}{0pt plus 0.3ex}
}

\title{Mass spectra of mesons containing charm quarks -- continuum
limit results from twisted mass fermions}

\ShortTitle{Mass spectra of mesons containing charm quarks from twisted mass LQCD}

\author{\speaker{Krzysztof Cichy}
\\ Goethe-Universit\"at Frankfurt am Main, Institut f\"ur Theoretische Physik,
Max-von-Laue-Str. 1, D-60438 Frankfurt am Main, Germany       \\
Adam Mickiewicz University, Faculty of Physics,
Umultowska 85, 61-614 Pozna\'n, Poland\\
        E-mail: \email{kcichy@th.physik-uni-frankfurt.de}}

\author{Martin Kalinowski\\
        Goethe-Universit\"at Frankfurt am Main, Institut f\"ur Theoretische 
Physik,
Max-von-Laue-Str. 1, D-60438 Frankfurt am Main, Germany       \\
        E-mail: \email{kalinowm@th.physik-uni-frankfurt.de}}

\author{Marc Wagner\\
        Goethe-Universit\"at Frankfurt am Main, Institut f\"ur Theoretische 
Physik,
Max-von-Laue-Str. 1, D-60438 Frankfurt am Main, Germany       \\
        E-mail: \email{mwagner@th.physik-uni-frankfurt.de}}

\abstract{We present results from an ongoing computation of masses of $D$ mesons, 
$D_s$ mesons and charmonium, including both ground states and several parity and 
angular momentum excitations. We employ 2+1+1 flavours of dynamical maximally 
twisted mass fermions at three lattice spacings and three $u/d$ quark masses at 
each lattice spacing. We consider different combinations of valence quark discretizations, with 
either identical or opposite signs in front of the twisted mass terms. In the 
end, our setup allows for a good control of different kinds of systematic 
effects, in particular the quark mass dependence of the resulting meson masses 
and cut-off effects. We obtain good agreement with experiment for the majority of states and we discuss improvements that will be made to finalize the analysis.
}

\FullConference{The 33rd International Symposium on Lattice Field Theory\\
		14 -18 July 2015\\
		Kobe International Conference Center, Kobe, Japan}

\begin{document}

\section{Introduction}
\vspace*{-0.2cm}
Meanwhile, there is a large number of mesons known experimentally which contain charm quarks. Some of them are well established and in good agreement with phenomenological expectations, but
in other cases their masses and/or widths are not well understood theoretically. For example, the $D_{s0}^*$ and $D_{s1}$ mesons are conjectured to be tetraquark candidates or mixtures of a mesonic and a tetraquark structure.
Hence, an \emph{ab initio} investigation of charmed mesons is highly interesting and can be in principle realized on the lattice.
However, charm physics on the lattice is complicated due to the currently feasible values of the lattice spacing -- if they are too coarse, the charm quark mass is large in lattice units. Nevertheless, with current computational resources many questions can be addressed, including the spectrum of charmed mesons.
Moreover, charm quarks can be treated as dynamical, so all systematic effects can be controlled with reasonable precision.
For recent lattice QCD papers using quark-antiquark operators for charm quark containing mesons, see e.g.\ Refs.\  \cite{Dong:2009wk,Burch:2009az,Dudek:2010wm,Mohler:2011ke,Namekawa:2011wt,Bali:2011dc,Bali:2011rd,Liu:2012ze,Yang:2012mya,Dowdall:2012ab,Moir:2013ub,Galloway:2014tta,Bali:2015lka}. 
For papers using additionally four-quark interpolating operators, cf.\ Refs.\  \cite{Gong:2011nr,Mohler:2012na,Liu:2012zya,Prelovsek:2013cra,Ikeda:2013vwa,Lang:2014yfa,Prelovsek:2014swa,Guerrieri:2014nxa}.
Our goal is to compute the spectrum of $D$ mesons (charm-light), $D_s$ mesons (charm-strange) and  charmonium (charm-charm)  using fully dynamical twisted mass ensembles generated by the European Twisted Mass Collaboration (ETMC) with 2+1+1 flavours. 
The results reported in this proceeding are from an ongoing work aiming at extending Ref.\ \cite{Kalinowski:2015bwa} to obtain the results in the continuum limit. 

\vspace*{-0.2cm}
\section{Lattice setup and lattice techniques}
\vspace*{-0.2cm}
We use dynamical twisted mass (TM) configurations generated by ETMC with 2+1+1 dynamical quark flavours \cite{Baron:2010bv}. The gauge action is the Iwasaki action \cite{Iwasaki:1996sn}, while the fermionic sector consists of the Wilson twisted mass action for the degenerate up/down doublet \cite{Frezzotti:2003ni} and non-degenerate strange/charm doublet \cite{Frezzotti:2003xj}. Automatic $\mathcal{O}(a)$ improvement is realized by setting the hopping parameter $\kappa$ to its critical value for which the PCAC quark mass vanishes \cite{Frezzotti:2003ni,Frezzotti:2004wz}.

In the valence sector, we use the following setup. The action for the light quarks is the same as the one in the sea.
For strange and charm, we introduce two strange ($s$, $s'$) and two charm ($c$, $c'$) quark flavours with the action for a single flavour $f$ \cite{Frezzotti:2004wz}:
\vspace*{-0.1cm}
\begin{equation}
D_f=D_W+m_0+i\mu_f\gamma_5. 
\vspace*{-0.25cm}
\end{equation}
We take
\vspace*{-0.1cm}
\begin{itemize}
\item either  $\mu_{s/c}=-\mu_{s'/c'}$  -- we call this TM setup (however, it is still non-unitary)
\vspace*{-0.2cm}
\item or $\mu_{s/c}=\mu_{s'/c'}$  -- we call this Osterwalder-Seiler (OS) setup.
\end{itemize}
In this way, we avoid the mixing of strange and charm quarks, which would make the computations problematic.
It is important to emphasize that such setup still guarantees automatic $\mathcal{O}(a)$ improvement.

\begin{table}[t!]
   \centering
  \begin{tabular}[]{cccccccccc}
    \multirow{2}{*}{Ensemble} & \multirow{2}{*}{$\beta$}& \multirow{2}{*}{lattice} &
\multirow{2}{*}{$a\mu_l$} & $\mu_{l,R}$&
    \multirow{2}{*}{$\kappa_c$}  & $L$  & \multirow{2}{*}{$m_\pi L$} & $a$ &
\\
&&&&[MeV]&&[fm]&&[fm]\\
\hline
A30.32 &1.90 & $32^3\times 64$  & 0.0030 & 13 & 0.163272  & 2.8 & 3.5 & 0.0885\\
A40.32 &1.90 & $32^3\times 64$  & 0.0040 & 17 & 0.163270  & 2.8 & 4.1 & 0.0885\\
A80.24 & 1.90 & $24^3\times 48$  & 0.0080 & 34 & 0.163260  & 2.1 & 4.3 & 0.0885 \\
B25.32 & 1.95 & $32^3\times64$  & 0.0025 & 12& 0.161240 & 2.6 & 3.2 & 0.0815 \\
B55.32 & 1.95 & $32^3\times64$  & 0.0055 &26& 0.161236 &2.6 & 4.6 & 0.0815 \\
D15.48 &  2.10 & $48^3\times96$  & 0.0015 & 9 &0.156361&3.0 & 3.2 & 0.0619 \\
D20.48 &  2.10 & $48^3\times96$  & 0.0020 & 12 &0.156357&3.0 & 3.7 & 0.0619 \\
D30.48 & 2.10 & $48^3\times96$  & 0.0030 & 19 &0.156355 &3.0 & 4.5 & 0.0619 \\
\end{tabular}
\caption{\label{tab:setup}Simulation parameters for gauge field configuration ensembles used in this work. Shown are: ensemble label, inverse gauge coupling ($\beta$), lattice volume $((L/a)^3\times(T/a))$, sea quark mass ($a\mu_l$), its physical value in MeV ($\mu_{l,R}$, renormalized in the $\MSb$ scheme at $\mu=2$ GeV), critical value of the hopping parameter yielding vanishing PCAC mass ($\kappa_c$), lattice extent in fm ($L$), product of the pion mass and the lattice extent ($m_\pi L$) and lattice spacing in fm ($a$).}
\end{table}

The simulation parameters of our ensembles are summarized in Tab.\ \ref{tab:setup}.
We use three lattice spacings between approximately 0.06 fm and 0.09 fm and pion masses ranging between around 230 MeV and 480 MeV.
This enables us to investigate the discretization and quark mass effects and extrapolate our results to the continuum limit and the physical pion mass.

Our lattice meson creation operators are of the following form\footnote{For a detailed account of the used lattice techniques, we refer to Ref.\ \cite{Kalinowski:2015bwa}.}:
\vspace*{-0.1cm}
\begin{equation}
O_{\Gamma,\bar{\chi}^{(1)} \chi^{(2)}}^{\scriptsize\textrm{tm}} \ \ \equiv \ \ \frac{1}{\sqrt{V/a^3}} \sum_\mathbf{n} \bar{\chi}^{(1)}(\mathbf{n}) \hspace*{-5mm}\sum_{\Delta \mathbf{n} = \pm \mathbf{e}_x , \pm \mathbf{e}_y , \pm \mathbf{e}_z} \hspace*{-8mm} U(\mathbf{n};\mathbf{n} + \Delta \mathbf{n}) \Gamma(\Delta \mathbf{n}) \chi^{(2)}(\mathbf{n} + \Delta \mathbf{n}), 
\end{equation}
where $\sum_\mathbf{n}$  gives zero total momentum, $\sum_{\Delta \mathbf{n}}$   realizes spatial separation between quarks (such that the meson can have orbital angular momentum), $\Gamma(\Delta \mathbf{n})$  is a suitable combination of spherical harmonics and $\gamma$-matrices (determines total angular momentum, parity and charge conjugation properties (for charmonium)), $U(\mathbf{n};\mathbf{n} + \Delta \mathbf{n})$  is a gauge link, ${\chi}^{(1),(2)}$ are twisted basis quark operators.

We use standard smearing techniques to enhance the overlap between trial states and low lying meson states. We apply APE smearing of links and Gaussian smearing of quark fields. 
Note that smearing does not affect the irreducible representation of the cubic group, equivalent of the total angular momentum $\mathrm{O}^J$ on the lattice, parity $\mathcal{P}$  and charge conjugation $\mathcal{C}$, all determined by $\Gamma(\Delta \mathbf{n})$ in the creation operators.

For each sector, i.e.\ the same flavours $\bar{\chi}^{(1)} \chi^{(2)}$, cubic representation $\mathrm{O}^J$  and (for $\bar{c}c$) $\mathcal{C}$  (OS) or $\mathcal{C} \circ \mathcal{P}^{(\textrm{tm})}$  (TM), we compute temporal correlation matrices of meson creation operators.
The different entries in a given correlation matrix hence differ by their $\Gamma$-structure (spin) and parity $\mathcal{P}$, since parity is broken by TM at finite lattice spacing.
An example of extraction of meson masses and assignment of parity from a $2\times2$ correlation matrix is discussed in detail in Ref.\ \cite{Kalinowski:2015bwa}, Sec.\ 4.2.

\begin{figure}[t!]
\begin{center}
\includegraphics[width=0.345\textwidth,angle=-90]{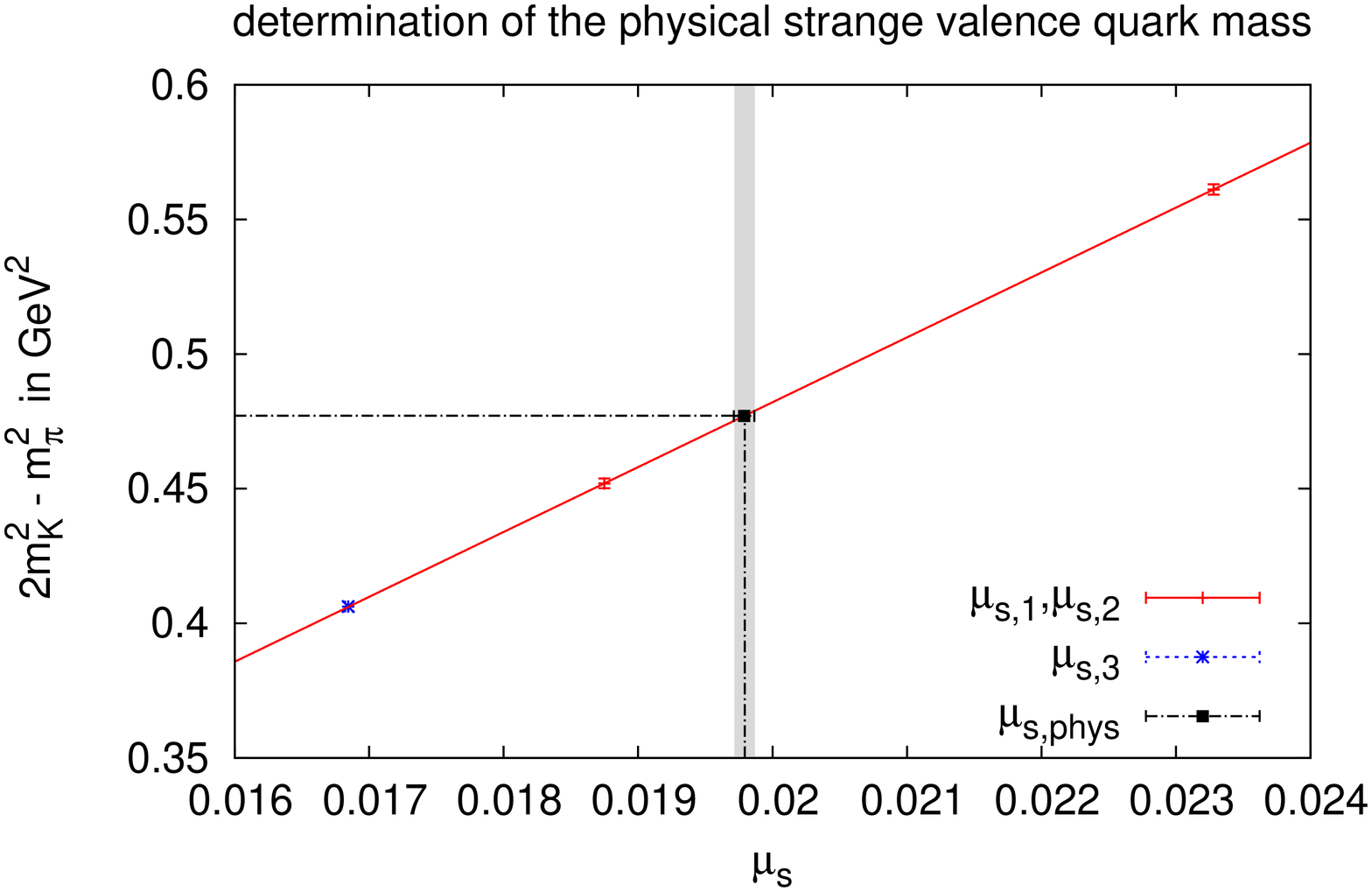}
\includegraphics[width=0.345\textwidth,angle=-90]{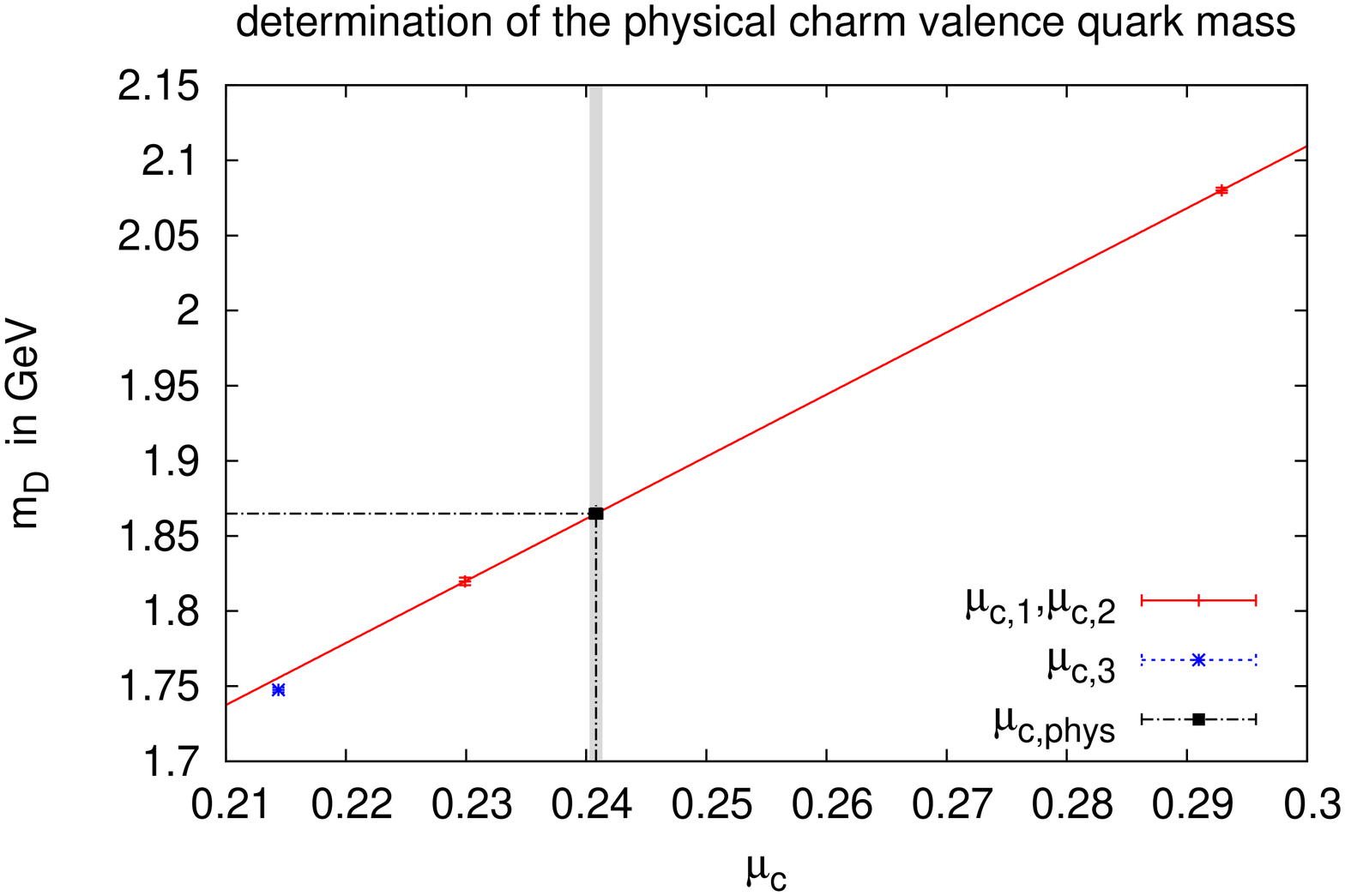}
\caption{\label{fig:tuning} Examples of tuning to physical strange (left) and charm (right) quark masses. Ensemble A80.24.} 
\end{center}
\end{figure}

\vspace*{-0.1cm}
\section{Results}
\vspace*{-0.25cm}
\subsection{Extrapolation procedure}
Our aim is to extract the physical masses of mesons containing charm quarks.
Therefore, we use the following procedure to extrapolate to the continuum limit and interpolate/extrapolate to the physical quark masses:
\vspace*{-0.2cm}
\begin{enumerate}
\item We compute the relevant TM/OS correlation functions for three lattice spacings, three light quark masses for each lattice spacing (with the exception of B-ensembles where the computations for one mass are still in progress), two strange quark masses per light quark mass and two charm quark masses per light quark mass (i.e.\ three pairs $(\mu_{s,1},\mu_{c,2})$, $(\mu_{s,2},\mu_{c,1})$, $(\mu_{s,2},\mu_{c,2})$ for each light quark mass  $\mu_l$).
\item We tune the strange/charm quark masses via $2m_K^2-m_\pi^2$ (which does not depend on the light quark mass at leading order of chiral perturbation theory) and the $D$ meson mass, $m_D$ (essentially light quark mass independent).
To compute $2m_K^2-m_\pi^2$ and $m_D$ for this tuning, we always use the TM setup ($\mu_{s,c}=-\mu_{s',c'}$).
The physical strange/charm quark masses are such values of $\mu_s/\mu_c$ that $2m_K^2-m_\pi^2$ and $m_D$ take their physical values of $0.477\,\textrm{GeV}^2$ and $1.865$ GeV, respectively (cf. Fig.\ \ref{fig:tuning}).
\item Using the values of the physical strange/charm quark masses $\mu_s/\mu_c$, we inter-/extrapolate all our meson masses.
We use jackknife with binning to account for autocorrelations and propagate the errors from the tuning.
\item This gives us a set of 16 points per meson mass (3 lattice spacings $\times$ (2--3) light quark masses $\times$ 2 discretizations).
Having this set of data points, we perform a combined chiral and continuum extrapolation, using the following fitting ans\"atze:
\vspace*{-0.15cm}
\begin{equation}
\label{eq:TM}
M^{TM}(a,m_\pi) =  M  +  c^{TM} a^2 +  \alpha^{TM}(m_\pi^2-m_{\pi,phys}^2), 
\vspace*{-0.1cm}
\end{equation}
\begin{equation}
\label{eq:OS}
M^{OS}(a,m_\pi) =  M  +  c^{OS} a^2 +  \alpha^{OS}(m_\pi^2-m_{\pi,phys}^2)
\end{equation}
with five fitting parameters: $M$, $c^{TM}$, $c^{OS}$, $\alpha^{TM}$, $\alpha^{OS}$.
Note that we \emph{enforce} a common continuum and physical pion mass limit $M$ for both discretizations.
\end{enumerate}

\begin{figure}[p!]
\begin{center}
\includegraphics[width=0.345\textwidth,angle=270]{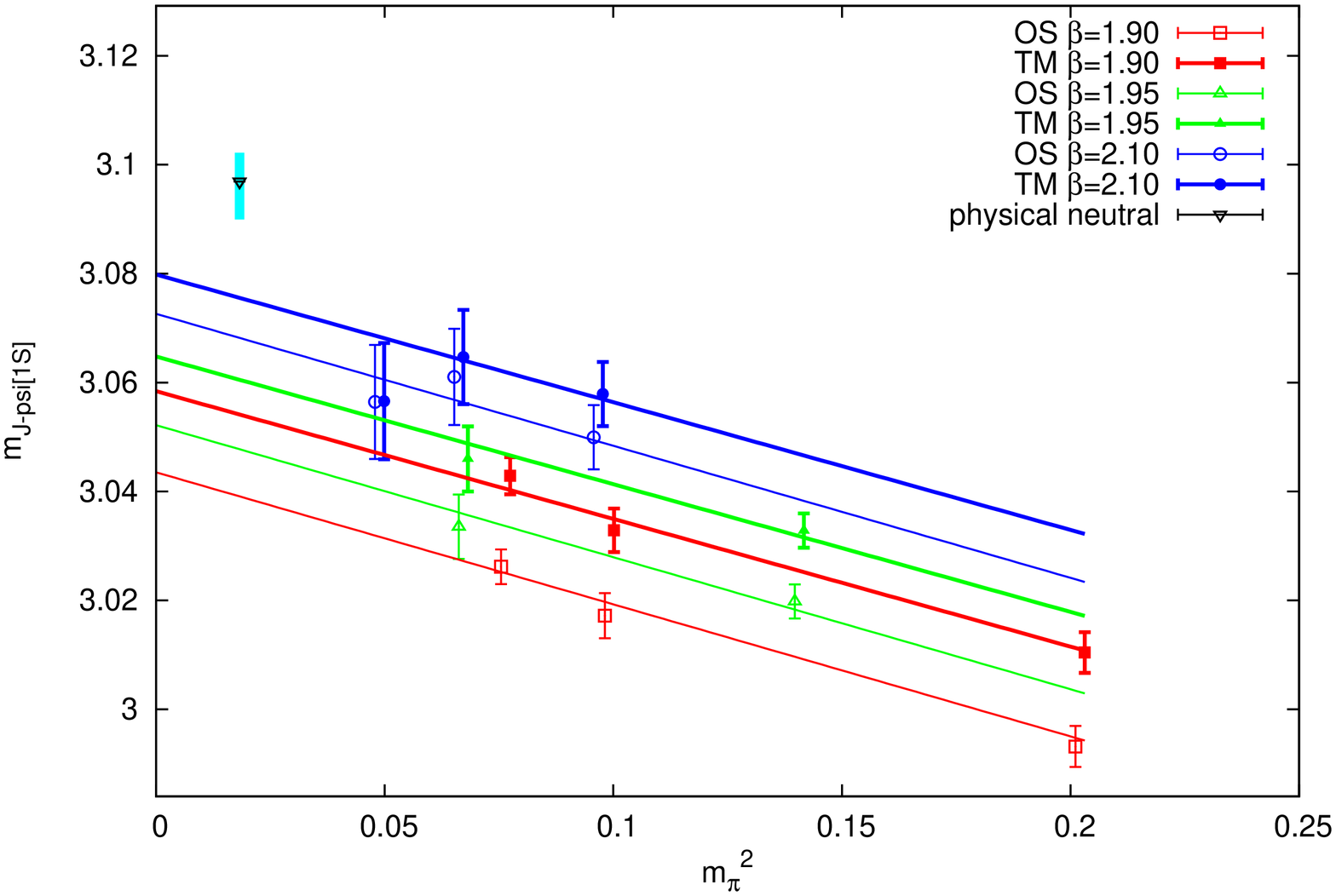}
\includegraphics[width=0.345\textwidth,angle=270]{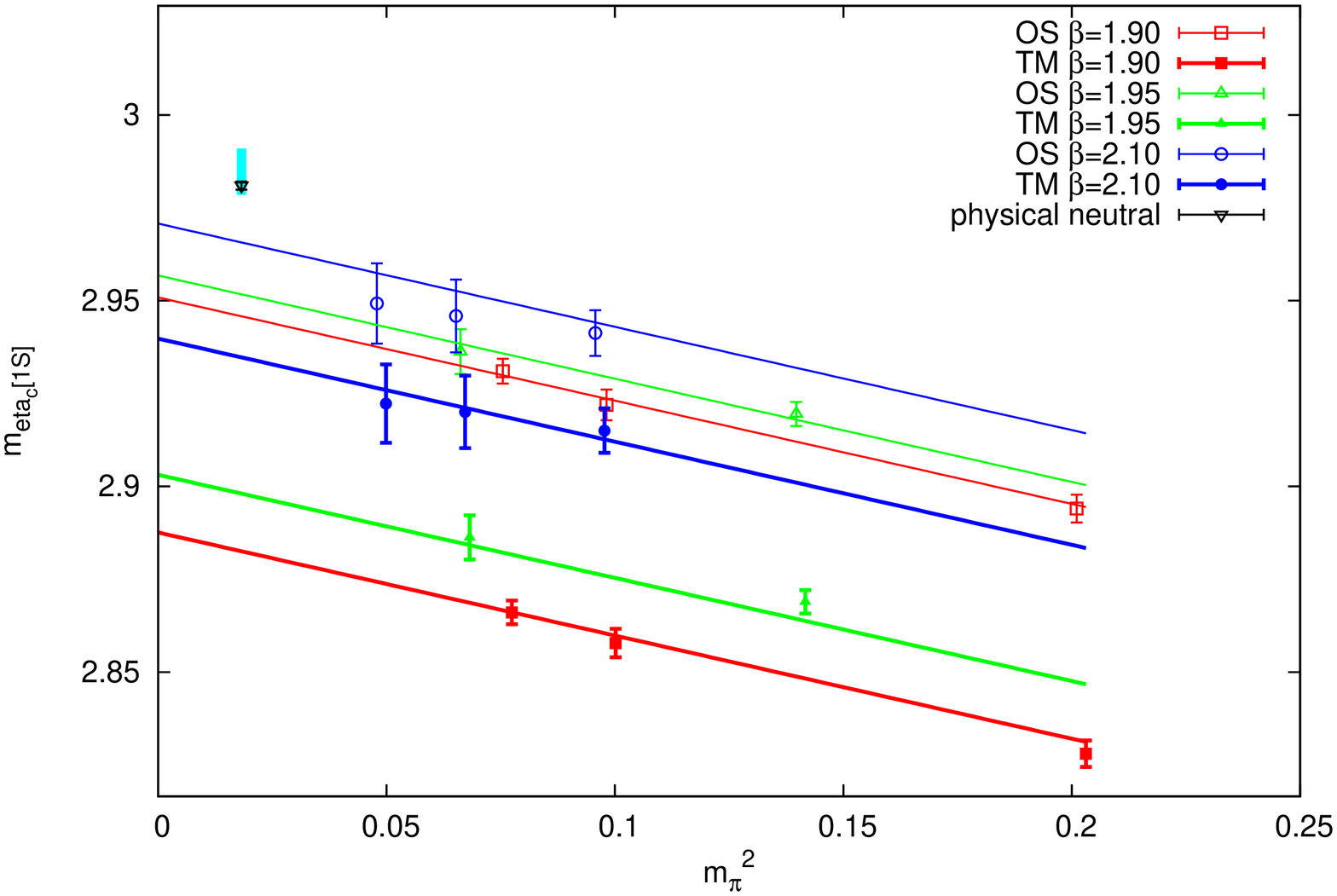}
\includegraphics[width=0.345\textwidth,angle=270]{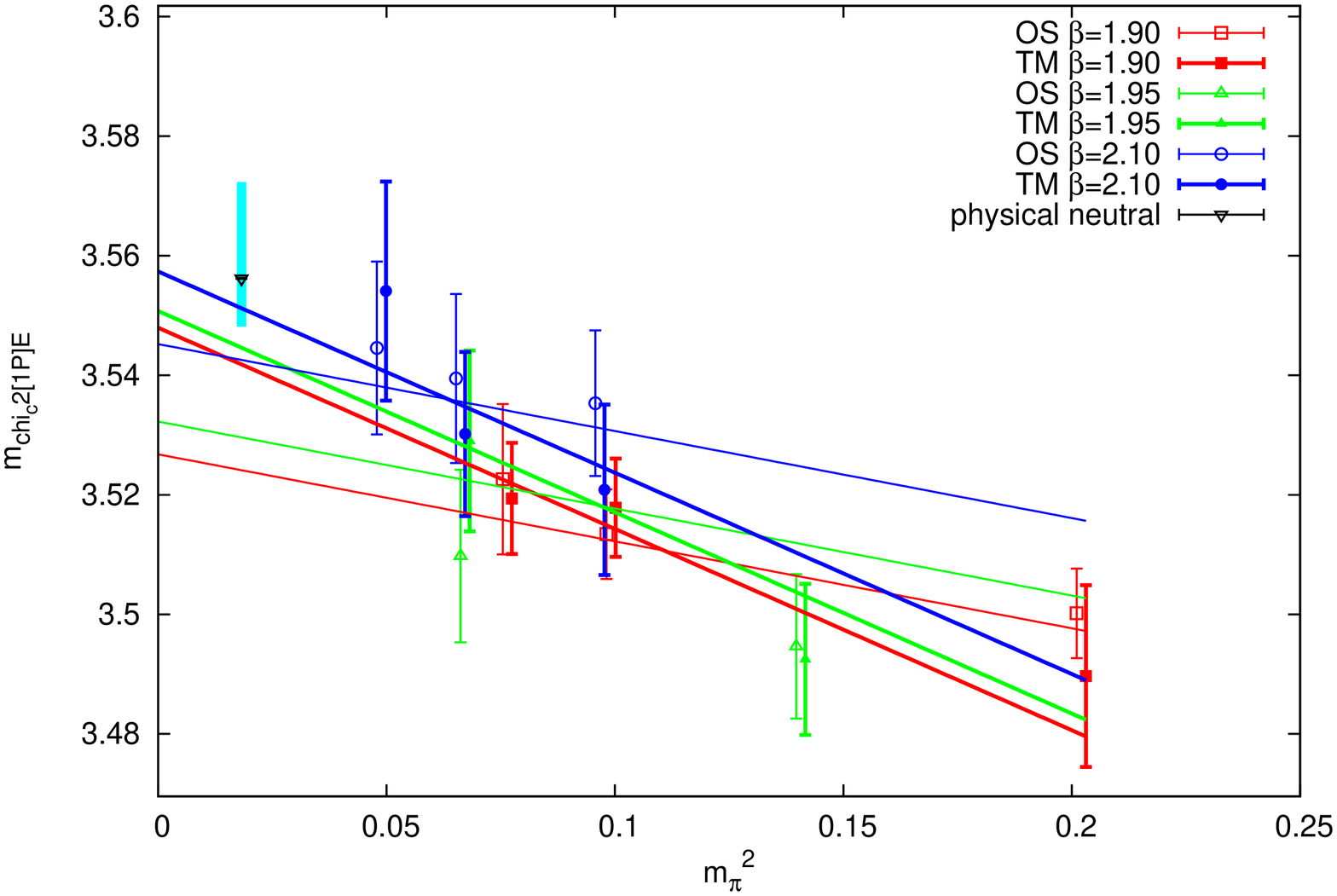}
\includegraphics[width=0.345\textwidth,angle=270]{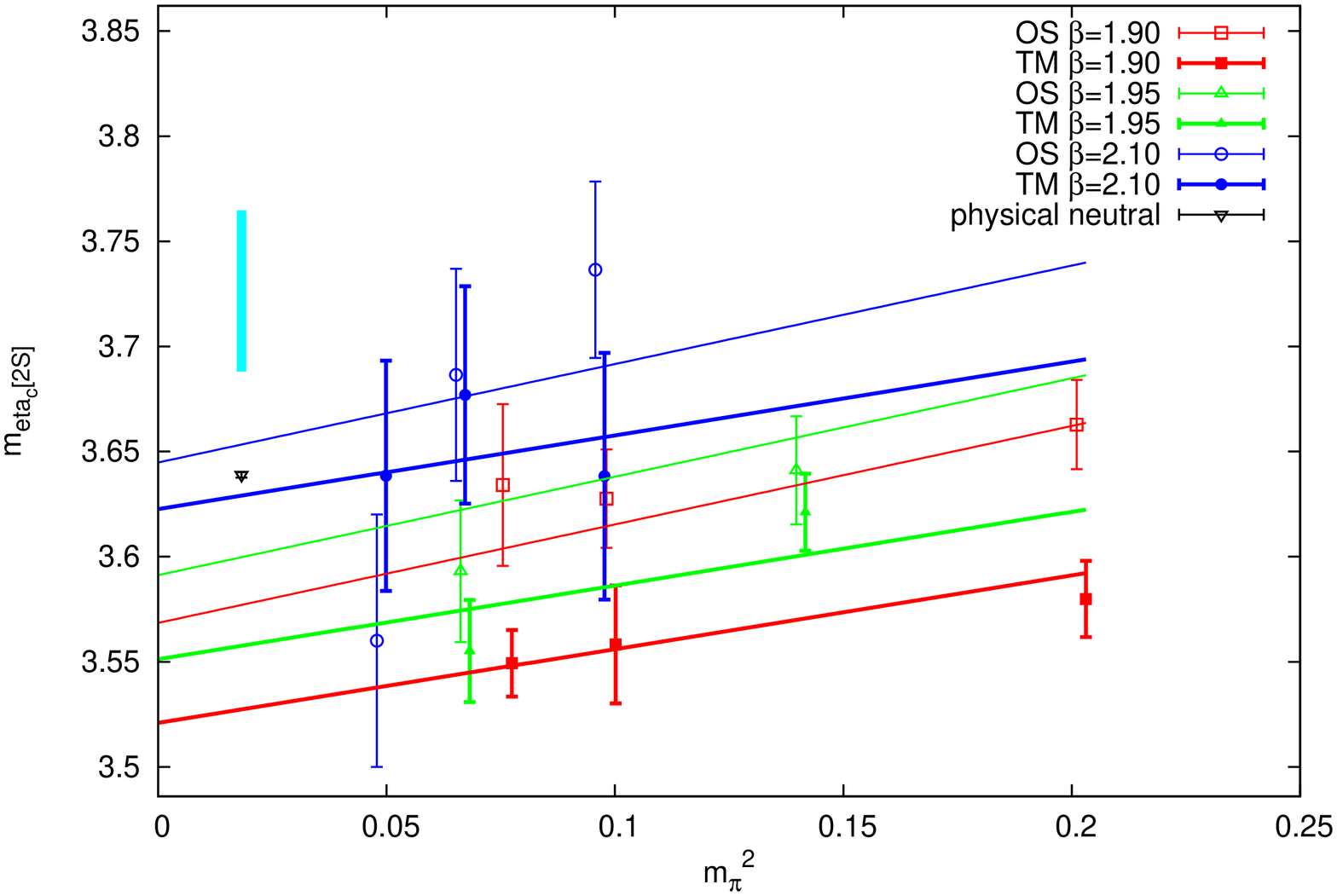}
\caption{\label{fig:cc}Combined chiral and continuum extrapolations in the charmonium sector: $J/\psi$ ($J^{\mathcal{P}\mathcal{C}}=1^{--}$, top left), $\eta_c(1S)$ ($J^{\mathcal{P}\mathcal{C}}=0^{-+}$, top right), $\chi_{c2}(1P)$ ($J^{\mathcal{P}\mathcal{C}}=2^{++}$, bottom left), $\eta_c(2S)$ ($J^{\mathcal{P}\mathcal{C}}=0^{-+}$, bottom right). PDG values of the masses \cite{PDG} vs. results of our extrapolations:  3096.920(10) MeV vs. 3096(6) MeV ($\chi^2/\textrm{d.o.f.}$ of our fit: 0.36), 2981.1(1.1) MeV vs. 2985(6) MeV ($\chi^2/\textrm{d.o.f.}=0.54$), 3556.20(9) MeV vs. 3560(12) MeV ($\chi^2/\textrm{d.o.f.}=0.53$), 3638.9(1.3) MeV vs. 3726(38) MeV ($\chi^2/\textrm{d.o.f.}=0.85$), respectively.} 
\end{center}
\begin{center}
\includegraphics[width=0.345\textwidth,angle=270]{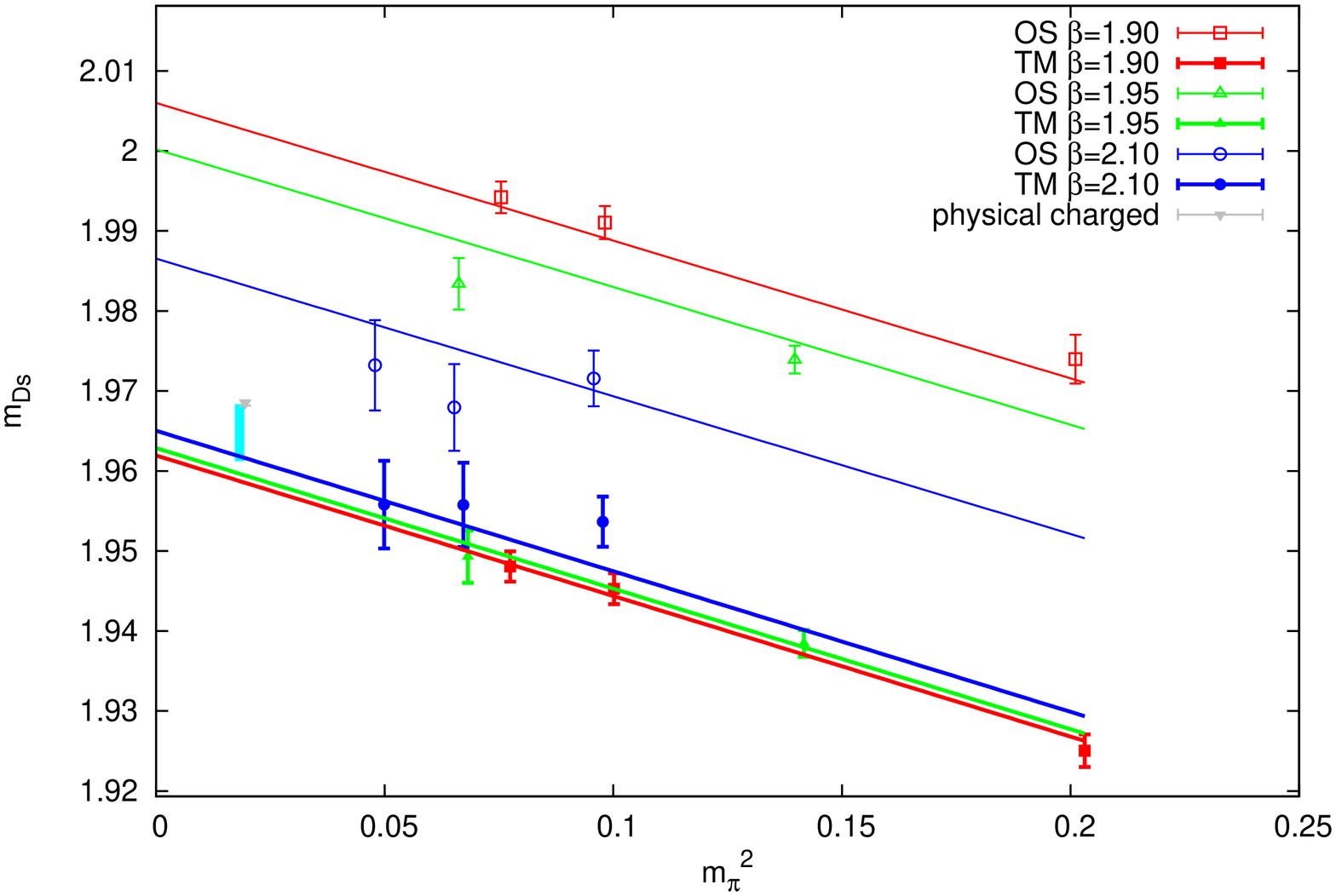}
\includegraphics[width=0.345\textwidth,angle=270]{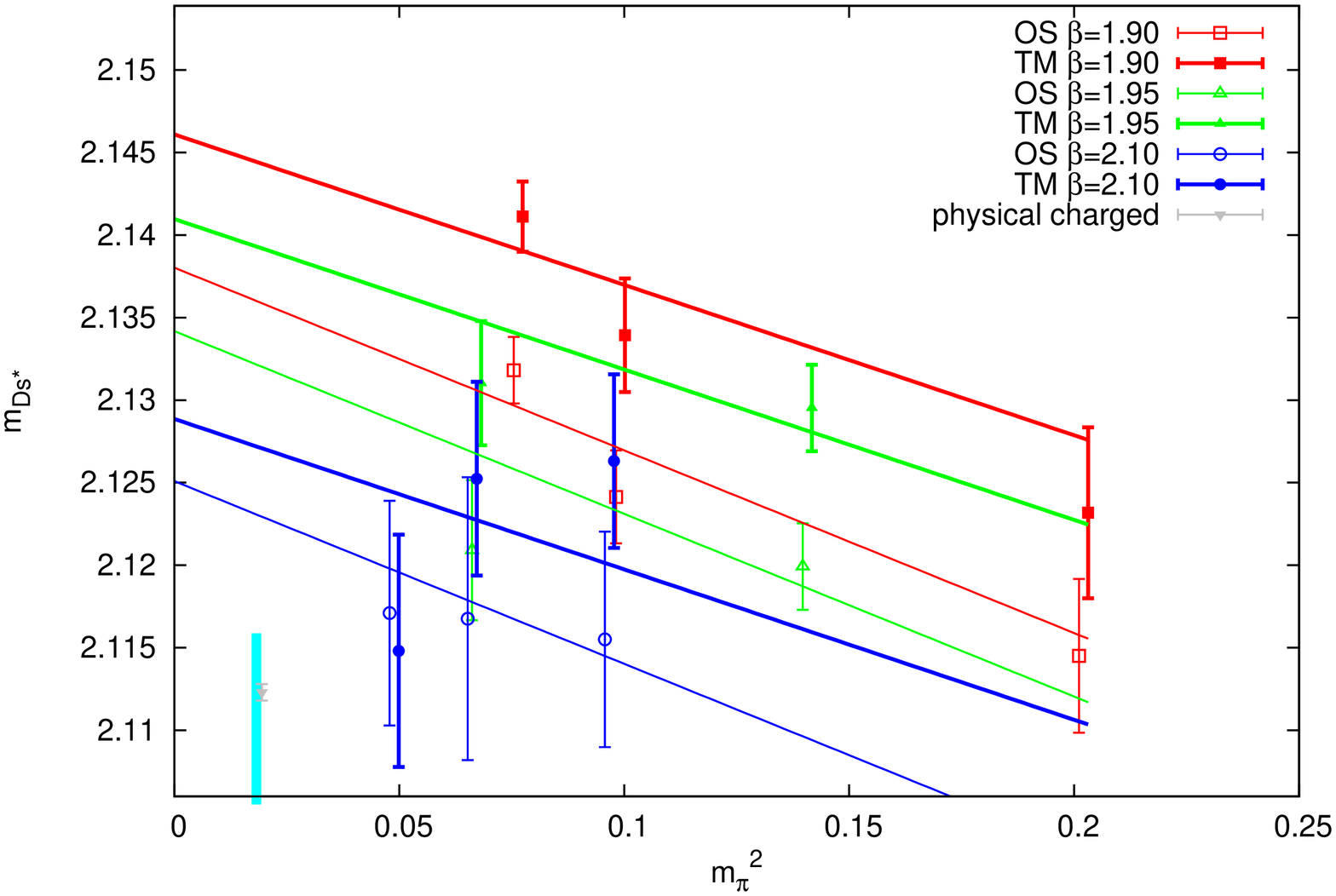}
\caption{\label{fig:Ds}Combined chiral and continuum extrapolations in the $D_s$ sector: $D_s$ ($J^{\mathcal{P}}=0^-$, left), $D_s^*$ ($J^{\mathcal{P}}=1^-$, right). PDG values of the masses \cite{PDG} vs. results of our extrapolations:  1968.49(32) MeV vs. 1964.8(3.6) MeV ($\chi^2/\textrm{d.o.f.}=1.24$), 2112.3(5) MeV vs. 2110.7(5.2) MeV ($\chi^2/\textrm{d.o.f.}=1.08$), respectively.} 
\end{center}
\end{figure} 

\vspace*{-0.3cm}
\subsection{Examples of combined chiral and continuum extrapolations and spectrum plots}
\vspace*{-0.1cm}
In Figs.\ \ref{fig:cc} and \ref{fig:Ds}, we present examples of our combined chiral and continuum extrapolations using Eqs.\ (\ref{eq:TM}) and (\ref{eq:OS}), four in the charmonium sector and two in the $D_s$ sector.
In all cases, our fitting ans\"atze give good description of lattice data and agreement with PDG values \cite{PDG} for the ground states of given channels ($J/\psi$, $\eta_c(1S)$, $\chi_{c2}(1P)$, $D_s$, $D_s^*$) at the per-mille level.
One of the presented cases ($\eta_c(2S)$) is an excited state (in the $J^{\mathcal{P}\mathcal{C}}=0^{-+}$ channel) and we observe discrepancy with respect to the PDG value.
We plan to investigate the sources of such discrepancies further -- in the cases of excited states, they are most probably due to short plateaus and hence a systematic analysis of uncertainties from the choice of the plateau fitting range is needed.

\begin{figure}[t!]
\begin{center}
\includegraphics[width=0.345\textwidth,angle=270]{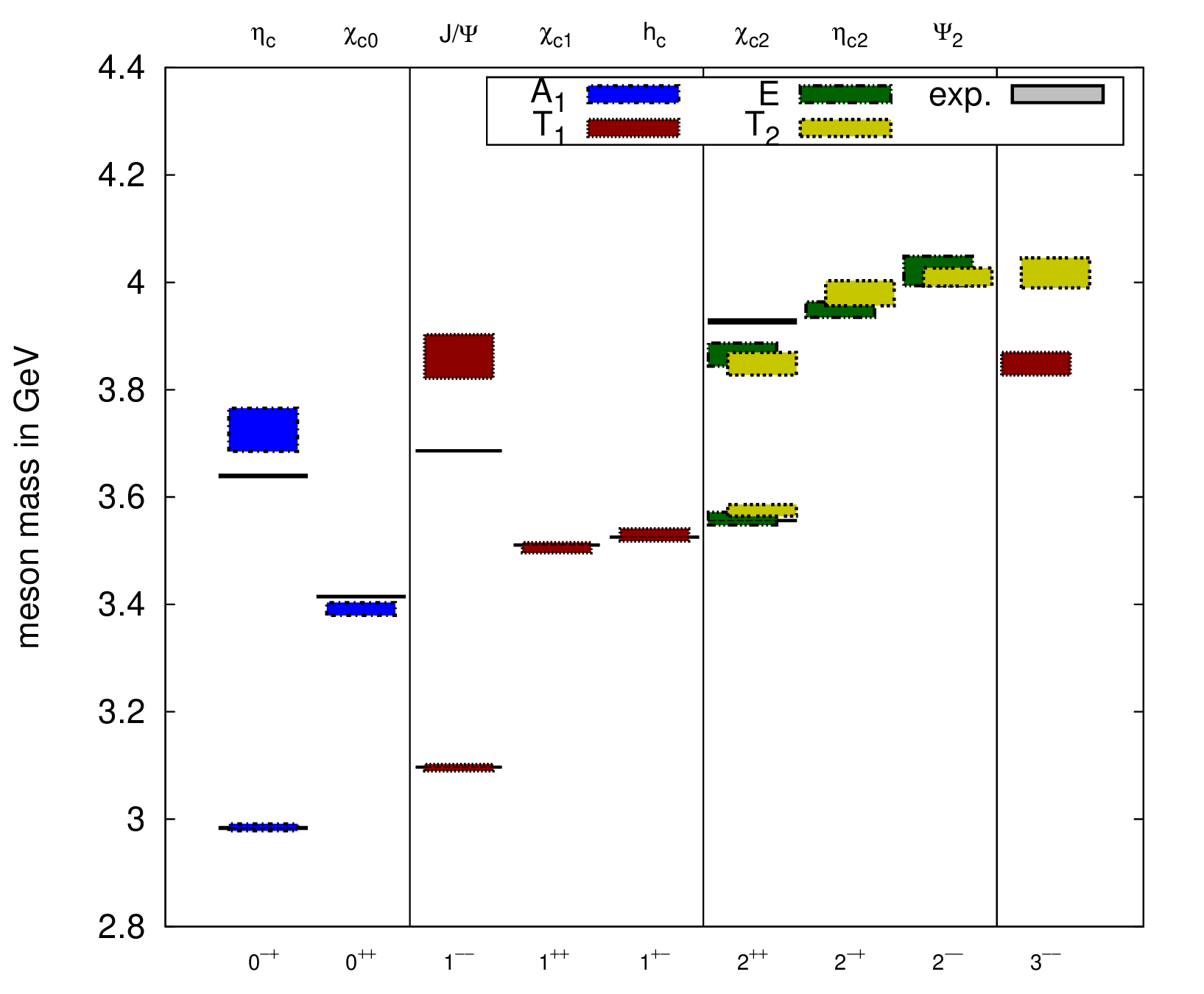}
\\
\includegraphics[width=0.345\textwidth,angle=270]{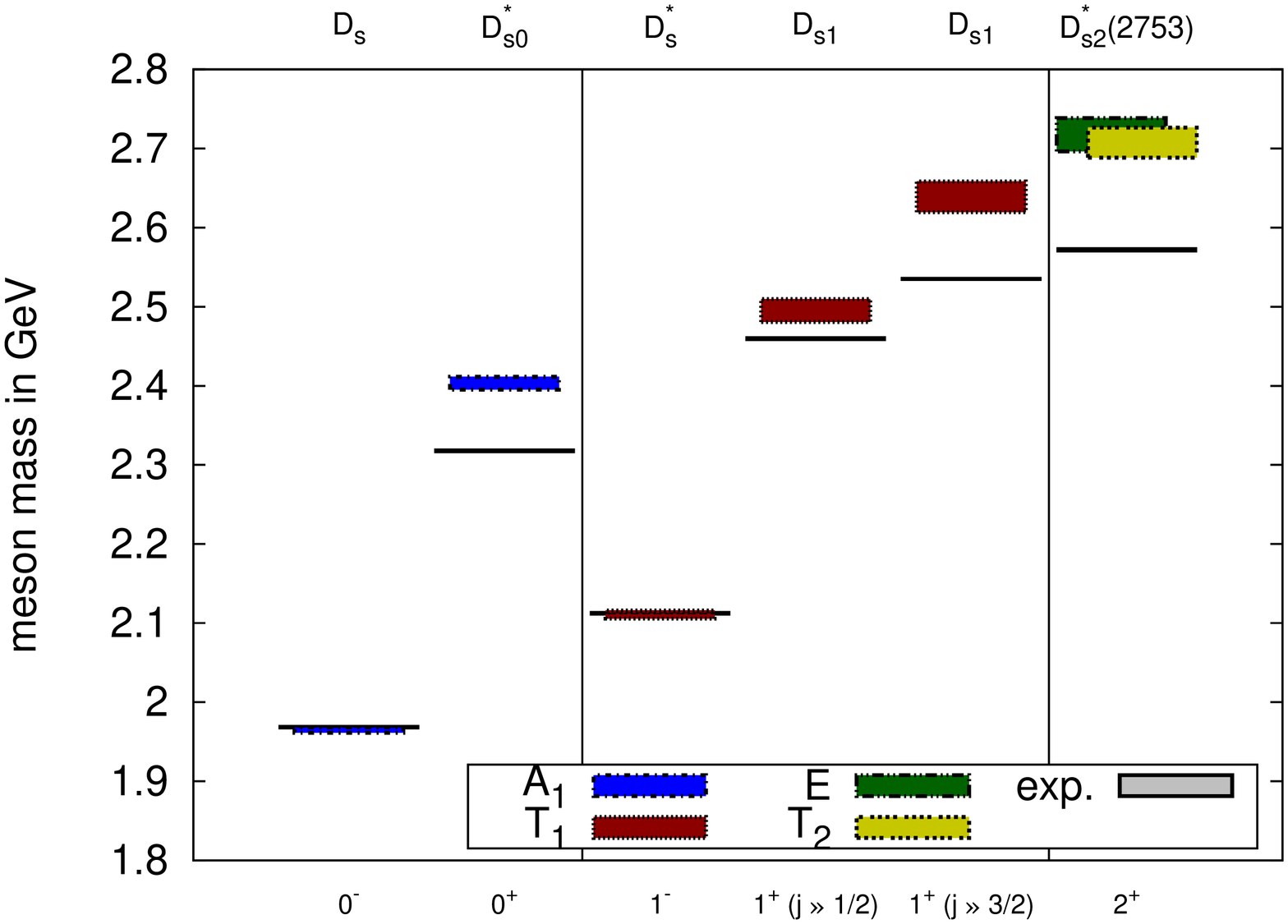}
\includegraphics[width=0.345\textwidth,angle=270]{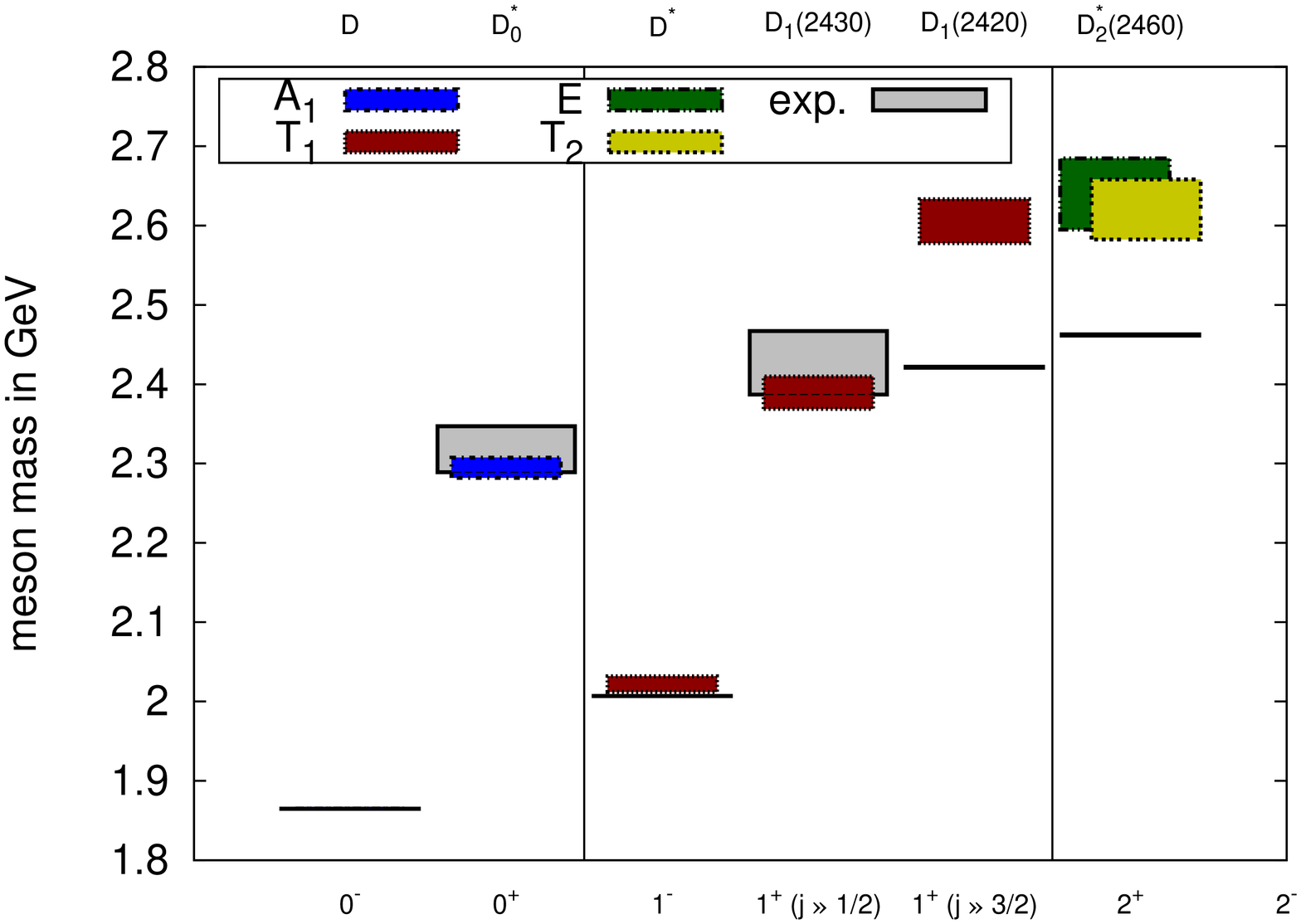}
\caption{\label{fig:spec} Spectrum plots for charmonium (top), $D_s$ mesons (bottom left) and $D$ mesons (bottom right). The black lines/grey boxes correspond to the PDG values \cite{PDG} (if available; the line/box widths correspond to the experimental uncertainties and/or resonance widths), while the coloured boxes are our lattice results in different representations of the cubic group (extrapolated to the continuum limit and the physical quark masses).}
\end{center}
\end{figure}

Finally, we present the summary of our results in Fig.\ \ref{fig:spec}, where we compare our lattice-extracted meson masses for charmonium, $D_s$ and $D$ mesons with PDG values \cite{PDG}.
Our results at this stage are still preliminary -- in the near future we will extend our analysis \cite{prep} by including the missing quark mass at one of the lattice spacings and by investigating in more detail the sources of systematic uncertainties, such as the ones related to the choice of the plateau range (particularly for excited states), finite volume effects and the choice of the fitting ansatz to extrapolate to the physical pion mass and the continuum limit.
Nevertheless, we already have rather good control over quark mass effects and discretization effects.
For many cases, we obtain good agreement with PDG values, especially for ground states, where the plateau quality is good.
Note that in certain cases we don't expect agreement with experiment, since we assume a dominating $q\bar{q}$ structure (whereas e.g.\ the $D_{s0}^*$ might be a tetraquark) and that the particles are stable (while e.g.\ the $D_0^*$ can decay to $D+\pi$ and hence should possibly be treated with more advanced lattice techniques).

\vspace*{-0.2cm}
\section*{Acknowledgments}
\vspace*{-0.25cm}
We acknowledge support by the Emmy Noether Programme of the DFG (German Research
Foundation), grant WA 3000/1-1.
This work was supported in part by the Helmholtz International Center for FAIR within the
framework of the LOEWE program launched by the State of Hesse.
Calculations on the LOEWE-CSC and on the on the FUCHS-CSC high-performance computer of
the Frankfurt University were conducted for this research. We would like to thank HPC-Hessen,
funded by the State Ministry of Higher Education, Research and the Arts, for programming
advice.

\end{document}